# μSR-dtected Soft mode toward a Possible Phase Transition in disordered spin gap system $(CH_3)_2CHNH_3$-$Cu(Cl_xBr_{1-x})_3$

*T. Goto[a], T. Suzuki[b], I. Watanabe[b], K. Kanada[a], T. Saito[a], A. Oosawa[a], H. Manaka[c]

[a]*Faculty of Science and Technology, Sophia University 7-1 Kioicho, Chiyodaku Tokyo 102-8554, Japan*

[b]*Advanced Meson Science Laboratory, Nishina Center, RIKEN, 2-1 Hirosawa, Wako, Saitama 351-0198, Japan*

[c]*Graduate School of Science and Engineering, Kagoshima University Korimoto, Kagoshima 890-0065, Japan*

**Abstract**

Measurements of macroscopic properties have indicated that the bond-disordered spin-gap system $(CH_3)_2CHNH_3$-$Cu(Cl_xBr_{1-x})_3$ is gapless when $x$ is between 0.44 and 0.87. Using muon spin relaxation to investigate microscopic properties of sample with $x$=0.35, we observed a dynamical spin fluctuation, whose characteristic frequency decreases with decreasing temperature, indicating a magnetic ground state.

*Keywords:* spin gap system, bond-disorder, NMR, μSR

## 1. Introduction

The bond-disordered quantum spin system $(CH_3)_2CHNH_3$-$Cu(Cl_xBr_{1-x})_3$ here abbreviated IPA-$Cu(Cl_xBr_{1-x})_3$ is a solid solution of two spin gap systems IPA-$CuCl_3$ with an energy gap of 14 K, and IPA-$CuBr_3$, with an energy gap of 98 K [1-3]. IPA-$CuCl_3$ had been reported to be an $S$=1/2 ferromagnetic-antiferromagnetic bond-alternating chain oriented along the $c$-axis[3], but recently inelastic neutron scattering experiments have revealed that it is a two-legged spin ladder system with strongly-coupled ferromagnetic rungs, oriented along the along the $a$-axis [1]. The system can therefore be considered an effective $S$=1 antiferromagnetic chain, and thus a composite Haldane system, expected to have a gapped singlet ground state [4,5,6,7]. IPA-$CuBr_3$ on the other hand, seems to be an $S$=1/2 antiferromagnetic-antiferromagnetic bond-alternating chain [2]. The subject of this paper is to investigate the ground state of the solid solution of these two systems.

The results of previous measurements of macroscopic quantities (uniform susceptibility and specific heat) have indicated that the system becomes gapless and shows a magnetic order at $T_N$ = 12–18K (depending on $x$) only when $x$ is between 0.44 and 0.87 [8,9,10,11]. A theoretical investigation utilizing the quantum Monte Carlo method, however, showed that IPA-$Cu(Cl_xBr_{1-x})_3$ with $x$<0.44 should be gapless though the occurrence of the magnetic order is critical [12].

We have used muon spin relaxation (μSR) to investigate the ground state of sample of IPA-$Cu(Cl_xBr_{1-x})_3$ single crystals with $x$=0.35 microscopically. As is reported on many other quantum spin systems[14,15], one cannot judge their ground state to be gapped or gapless simply from the temperature dependence of the muon spin relaxation rate. In this article, we show that the characteristic frequency of dynamical spin fluctuation decreases toward zero with decreasing temperature, suggesting the existence of a magnetic order at absolute zero or an extremely low temperature.

---

* Corresponding author. Tel.: +81-3-3238-3356; e-mail: gotoo-t@@sophia.ac.jp

## 2. Experimental

Single crystals of IPA-Cu($Cl_xBr_{1-x}$)$_3$ with $x$=0.35 were prepared by an evaporation method [2,3,8,9]. The concentration of Cl in the crystals was determined by inductively coupled plasma atomic emission spectrometry. Measurements of μSR were carried out at the Riken-RAL Muon Facility in the U. K. using a spin-polarized pulsed surface-muon ($\mu^+$) with a momentum of 27 MeV/c. The decay of the muon spin polarization is described by the time evolution of the asymmetry parameter $A(t)$, which is proportional to the spin polarization of the muon ensemble and is obtained from the ratio of the numbers of muon-events counted by the forward and backward counters.

The aggregations of aligned single crystals (total volume ≈400 mm$^3$) used in the μSR experiments were attached with Apiezon N grease to a 99.999%-pure silver plate connected to the bottom of the cryostat. The incident muon beam was parallel to the $b^*$-axis, which was perpendicular to the C-plane (one of the three outer planes with the largest area) [3].

## 3. Results and discussions

Typical relaxation curves of the muon spin polarization decay observed under various longitudinal fields (LFs) are shown in Fig. 1. No muon spin rotation due to a static field produced by a magnetically ordered phase was observed at any temperature down to 0.3 K or under any magnetic field, which is consistent with the measured macroscopic properties of IPA-Cu($Cl_xBr_{1-x}$)$_3$ [8,9]. The muon spin relaxation was described by the same function as adopted in our previous report[13], containing two components expressed as

$A_1 G_{KT}(t,\Delta)\exp(-t/\lambda_1) + A_2\exp(-t/\lambda_2)$

where $G_{KT}$ is the Kubo-Toyabe function [18,19] with parameters of a static field distribution width at the muon site $\Delta$, relaxation rates $\lambda_1$ and $\lambda_2$, and component amplitudes $A_1$ and $A_2$. The static field distribution width $\Delta$ was about 0.2 μs$^{-1}$, was temperature independent within an experimental resolution, and was considered to be from the quasi static nuclear spin contribution [18,19,14]. Note that the magnitude of $\lambda_2$ is very large, hiding the effect of nuclear contribution, so that $G_{KT}$ is omitted in the second term.

The amplitude fraction of $\lambda_2$, the faster relaxation, was around $A_2/(A_1+A_2) \approx 0.1$ at 8K, increased with decreasing temperature, and reached 0.4 at 0.3 K. The change in the $A_2$ with temperature is simply a consequence of the fact that the present system is very close to the quantum critical point $x$=0.44.

Generally, the LF-dependence of $\lambda$ gives a frequency spectrum of a dynamically fluctuating spins. At the lowest temperature 0.3K, the LF-dependences of the two relaxation rates obey the Redfield equation and correspond to the behaviour of classical localized spins fluctuating paramagnetically with a single characteristic time constant [25]. By fitting the observed $\lambda(H_{LF})$ to the function [25] expressed as

$\lambda(H_{LF})=2(\gamma_\mu \delta H_{loc})^2\tau_C/(1+\gamma_\mu H_{LF}\tau_C)^2$

where $\delta H_{loc}$, $\tau_c$ and $\gamma_\mu$ are respectively the local-field fluctuation amplitude, the characteristic fluctuation time constant, and the muon spin gyro-magnetic ratio (13.5534 kHz/Oe), we determined $\delta H_{loc}$ and $\tau_c$ to be 7.0±0.6 Oe and 7.6±0.5 μs for the $A_1$ component, and 38.5±4.0 Oe and 1.8±0.2 μs for the $A_2$ component.

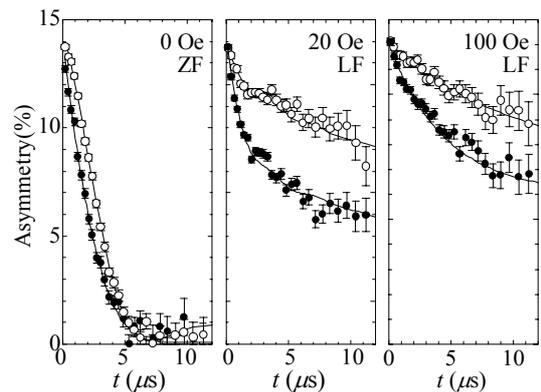

Fig.1 Typical muon polarization decay at (○) 0.3 K and (●) 5 K under various longitudinal fields. Dashed curves correspond to the fitting equation described in the text.

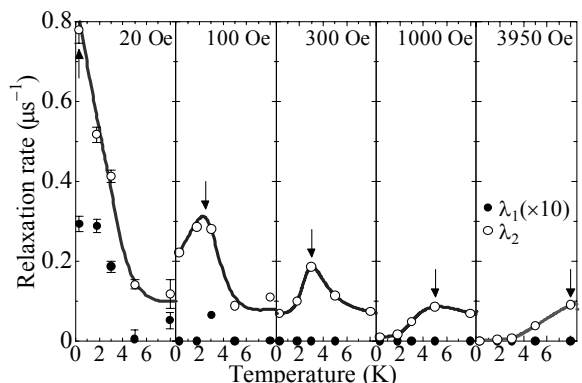

Fig. 2 Temperature dependence of the two relaxation rates $\lambda_1$ and $\lambda_2$ at various LFs. Arrows show the maximum $\lambda_2$ at each LF.

The appearance of two relaxation rates with different characteristic time constants indicates that the sample is a microscopically phase separated system. That is, it contains microscopic magnetic islands corresponding to $\lambda_2$ that are surrounded by a and a singlet-like sea corresponding to $\lambda_1$. This microscopic phase separation has also been found in Cl-rich mixtures ($x>0.87$) [13,10,16].

A possibility of macroscopic phase separation is readily ruled out, because the system does not show any magnetic order at low temperature down to 0.3 K assuring that the system does not bear any macroscopic part with concentration belonging to the magnetic region $0.44<x<0.87$.

The spin fluctuation frequency of $\lambda_1$ is significantly slower than that of the end member compound IPA-$CuCl_3$, which was reported to have a $\tau_C \approx 0.01$ μs [27]. This slow down is considered as an *exudation* of the antiferromagnetic spin correlation within magnetic islands to the singlet-sea. The fact that the spin fluctuation spectra of both the two regions are described by the Redfield equation assures that the system is still in the paramagnetic state at 0.3 K.

As the temperature is raised from 0.3 K, $\lambda(H_{LF})$ is not described by the simple Redfield equation, indicating that the fluctuation at higher temperatures cannot be described with a single time constant at high temperatures. We therefore instead the temperature dependence of $\lambda_1$ and $\lambda_2$ under various LF's as shown in Fig. 2. Note that these types of measurements, evaluating λ over wide ranges of temperature and LF, can be made only when using a high intensity muon beam like the ISIS-muon source at the Rutherford Appleton Laboratory.

Focusing on $\lambda_2$, we see that the spectrum weight at high frequency, which corresponds to high LF, decreases with decreasing temperature and that the weight at low frequency, which corresponds to low LF, increases. Tracing the equal temperature points in Fig. 2, one can also see that the spectrum is nearly white at high temperature, and heaps up around the low frequency region at low temperatures, reflecting the freezing behaviour of spin fluctuation. With decreasing temperature, $\lambda_2$ shows a peak structure at a temperature where the characteristic fluctuation frequency is thought to coincide with $\nu = \gamma_\mu H_{LF}$. We plot this temperature as a function of $\gamma_\mu H_{LF}$ in Fig. 3.

Although the error bars in Fig. 3 are quite large, one can clearly see that the characteristic frequency of the spin fluctuation decreases with decreasing temperatures. This freezing behaviour is interpreted as evidence of a soft-mode preceding a possible magnetic phase transition at absolute zero such as a Bose-glass phase. This exotic phase was first predicted theoretically by Fisher *et al.*[26] as a ground state of disordered boson systems in which particles are in gapless state and are still localized. The boson particles and vacuum in Fisher's notation are mapped onto the spin triplet and singlet sites in the present system. Although both macroscopic and microscopic techniques have provided evidence of a Bose-glass phase in the (Tl,K)$CuCl_3$ quantum spin system[20-24], the reality of such a phase is still in question.

The present result supports the existence of the

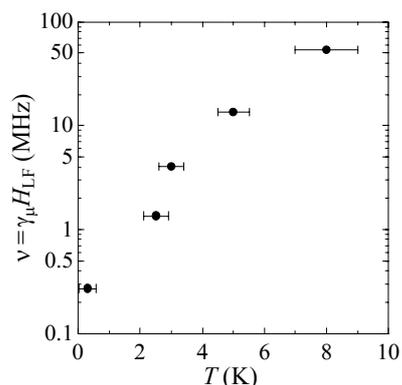

Fig.3 Temperature dependence of the characteristic frequency of the spin fluctuation in IPA-$Cu(Cl_{0.35}Br_{0.65})_3$.

Bose-glass phase, and brings together the two previous reports contradicting each other[8,12]. Since the critical temperature of the Bose-glass phase is absolute zero, one cannot observe directly the phase at finite temperature. This is the reason why the system was judged to be gapped in reports of macroscopic measurements [8]. The QMC calculation [12] concluded that the occurrence of the magnetic order is critical. This also reflects the fact that the critical temperature is zero. In the present study we succeeded to glimpse the phase, because we detect the critical phenomena which appears in a finite temperature region adjacent with the critical temperature.

The temperature dependence of the characteristic frequency shown in Fig. 3 is not the power-law type function as seen in conventional phase transitions, which has the form $(1-T/T_C)^z$, where $z$ is a dynamical critical exponent. This is because the critical temperature of the Bose-glass is absolute zero, so the functional is not necessarily of power-law type.

Finally we note that we cannot deny the other possibilities such as an exotic ground state still other than Bose-glass phase or a phase transition with finite critical temperature much lower than $^3$He region. In order to confirm that the ground state of the present system is a Bose-glass, we need to investigate further the field-induced magnetically ordered phase, which is frequently called as the Bose-Einstein condensation (BEC) of magnons [28,17], to see whether or not the phase boundary curve between the Bose-glass phase and the BEC shows the characteristic curvature predicted by Fisher *et al*.[26]   This investigation is in progress.

**Acknowledgements**

Authors are grateful to Prof. Tomi Ohtsuki and Prof. Tota Nakamura for valuable discussion. This work was supported by KEK-MSL Inter-University Program for Overseas Muon Facilities and by a Grant-in-Aid for Scientific Research on the Priority Area "High Magnetic Field Spin Science in 100T" from MEXT.